# TRACE: A Time-Relational Approximate Cubing Engine for Fast Data Insights

**Suharsh Sivakumar, Jonathan Shen, Rajat Monga**

Inference.io

## Abstract

A large class of data questions can be modeled as identifying important *slices* of data driven by user defined metrics. This paper presents TRACE, a Time-Relational Approximate Cubing Engine that enables interactive analysis on such *slices* with a low upfront cost - both in space and computation. It does this by materializing the most important parts of the cube over time enabling interactive querying for a large class of analytical queries e.g. what part of my business has the highest revenue growth ([SubCategory=Sports Equipment, Gender=Female]), what slices are lagging in revenue per user ([State=CA, Age=20-30]). Many user defined metrics are supported including common aggregations such as SUM, COUNT, DISTINCT COUNT and more complex ones such as AVERAGE. We implemented and deployed TRACE for a variety of business use cases.

## Introduction

With the ever increasing data volume and dimensionality, analyzing data keeps getting harder and slower. The complexity of the space leads to a higher need for ad hoc exploration as the results of each successive query are used to decide what to explore next.

Unlike relational queries that the current class of databases and warehouses are well optimized for, we consider analytical questions that are modeled as identification of important slices of data. For example, what is driving the drop in conversion rate this week?

Answering such business questions requires considering all slices of data, and it is hard to specify them as relational queries and run on standard tables. More recently there is a body of work making it easier to formulate these analytical queries (HoCA, DIFF), however the runtime remains a bottleneck for interactive analysis.

Relational tables[1] allow answering of these questions for a single slice at a time, or limited to what "Group By" offers - comparing all slices for a given Attribute in the table e.g. all [Category=*]. What is needed here is comparison across all possible slices across all Attributes e.g. is [Category=Shoes] important or [Subscription=No] or the intersection of the two [Category=Shoes, Subscription=No]. Holistic Cube Analysis (HoCA) offers a way to write these types of queries, however each query needs to go through the entire dataset and explore all possible slices every time.

In this paper we provide a way to materialize this cube in a way that it can answer many of these questions, at an extremely low cost and providing an interactive analysis.

## Example Use cases

### Analytics today

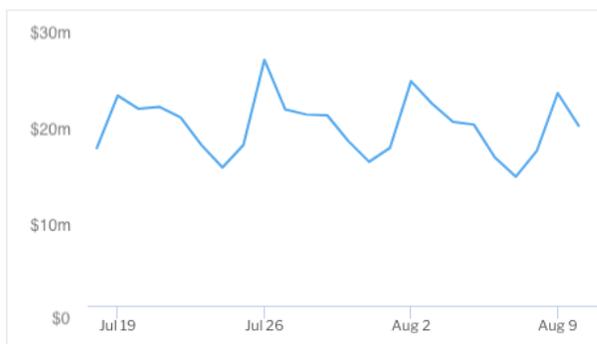

1. An E-commerce company tracks KPIs like Revenue & Users
2. User notices revenue seems down from the chart. Why is it?
3. Exploring 2-3 of 30+ dashboards, nothing seems obvious
4. The analyst can go through all the data. In reality, they look at the last few issues: [Channel=Email] or [Channel=Search] or [State=Texas] to see if it's the same issue.
5. The exploration is too slow and he gives up after a few hours hoping that the problem is gone the next day.

---

[1] Here we use *table* as a placeholder for relational tables, views, joins that may or may not be materialized. In fact the raw data doesn't even need to exist in a relational database for the purposes of our work.

Analytics with TRACE

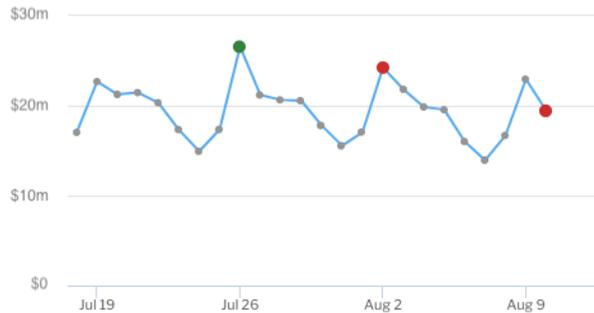

1. An anomaly detection algorithm uses TRACE to detect a drop in revenue and alert the user.
2. A relevance algorithm leverages TRACE to identify candidate segments and pinpoints the likely segment driving the issue (Kids Shoes in NY). All without the user having to ask.

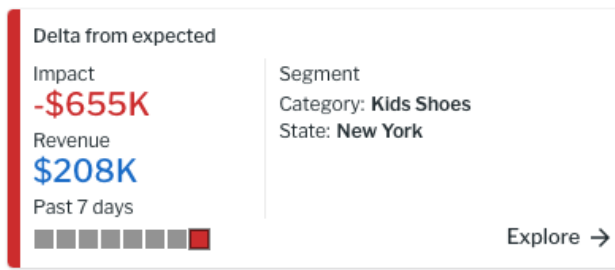

3. With additional data on other metrics it even points out that user views are driving the revenue down, and those are down because of a drop in users from FB.
4. The user can easily explore these metrics and get results like these interactively.

Almost every metric driven use case can benefit from TRACE. Here are some example metrics:

## Ecommerce

**Revenue** or Total Sales.
**Average Order Value** as Revenue over Number of Orders across various slices.

## Marketing

**Customer Acquisition Cost** as the total acquisition costs over the number of customers acquired. Lower is better.
**Customer Lifetime Value** typically per customer across the various slices. Higher is better. Helps drive investment in the most valuable slices.

### Product

**Daily Active Users** as the unique number of users on the product every day.
**Conversion Rate** as the total conversions (typically sales) over the number of incoming users.

## Definitions

### Metric

A user defined aggregation across a *Field* in a Table. For numeric aggregations such as SUM and AVERAGE, the field needs to be of type Number, while for COUNT or DISTINCT COUNT it may be any *Field*. Note: This works for metrics where higher values are more important, and values near zero are less important. This is true for most common metrics such as the ones discussed. It is ok for values to be negative, just that the more negative values are considered more important than near zeros in that case.

### Attribute

A *Field* in a Table that a metric may be sliced by e.g. Category, State or even Age Range (<20, 20-30, …).

### Slice

A specific combination of Attribute Values e.g. [Category=Shoes] is a slice with all rows that have Category=Shoes, leaving all other Attributes with any value, [State=CA] specifies value for State, and [Category=Shoes, (and) State=CA] specifies value for both Category and State, leaving all other Attributes with any value. The empty set [] (or top level) is also treated as a slice.

## Innovation

Cube materialization cost grows exponentially with the increase in dimensionality and with the increasing complexity in recent years cubes have fallen out of favor. The expense for any real world datasets has led to current approaches that run every query against the full dataset. Our cubing approach presented here materializes a cube for the computational cost of a single analysis (This is equivalent to 1 HoCA query since Relational Queries don't cover what's needed), for an extremely small storage cost. We'll show how once materialized, it reduces the cost for a variety of advanced analyses for a dataset with N rows and d columns from $O(N \times 2^d)$ to $O(n)$ with n as a tunable hyperparameter typically in the range 100k-1M, converting a runtime of minutes or hours to milli-seconds making it useful in interactive contexts.

For a given metric e.g. Revenue, materializing aggregations (SUM) for all slices for a fixed time granularity e.g. a day would typically take $O(N \times 2^d)$ both in storage and compute[2]. Instead, we allow for approximations using sketching techniques ([Charikar](#), [datasketches.org](#)) that enable us to pick the top n slices with a single pass over the data. It is these top n slices that are stored, for each time period (considering time as an additional dimension) leading to an $O(n)$ storage cost, and a one time $O(N \times 2^d)$ computation cost, the minimum required for the first such query on the raw data. The cube query in this case takes O(n) and achieves an extremely high recall, allowing them to be interactive (milliseconds) vs minutes in traditional approaches.

## Algorithm

The algorithm leverages ideas from streaming algorithms that leverage sketching techniques to get approximate results in a single pass. We start with the simple case of a single metric e.g. COUNT or SUM of a column in the record. For each record, generate all possible slices the record may be a part of, and for each of these slices, increment it with the value of the metric for that record e.g. for SUM this would be the value of the column, for COUNT by 1. These slices are typically maintained in a hash table for quick access. Since we wish to limit the number of slices we keep track of, we have to periodically prune this list to ensure that only the Top n are stored. However once the list is pruned, any new entries can no longer be guaranteed to not have been seen before. Hence we also keep track of the maximum value that was pruned out ($pruned_{max}$), and any new entry added after the pruning is initialized with the range [0, $pruned_{max}$] (Similar to AMC in [Macrobase](#)).

A large class of queries care about these metrics in specific time periods e.g. week, day, hour. In addition to keeping these sketches for the entirety of the data, we additionally keep sketches for each new time period e.g. every day, week and month that are usual for business metrics. Note, since data typically comes in chronological order, typically only the newest time periods need to be updated. Bigger time periods can be maintained directly, or alternatively generated by combining sketches for smaller time periods at the cost of some approximation. For applications such as IT monitoring, the time periods could be as small as every minute.

---

[2] The $2^d$ comes from exploring all dimension combinations from C(d, 1) to C(d, d). In practice it is lower because it is not meaningful to go beyond the combinations of a few dimensions.

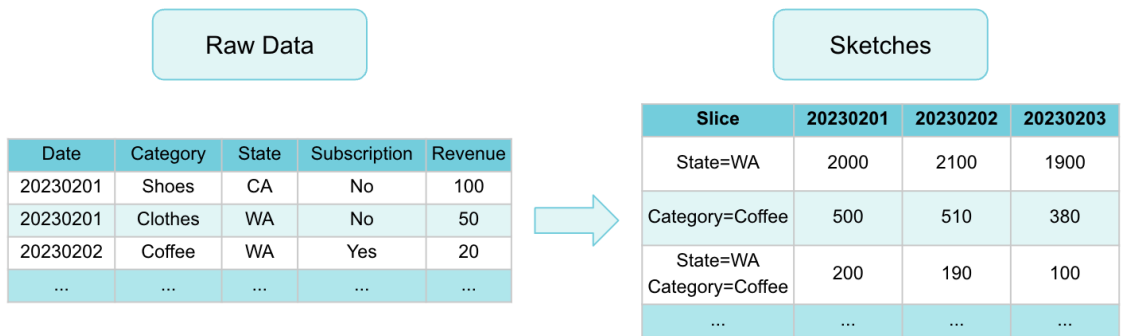

Once the Top n slices are available for each time period for a metric such as Revenue, this allows building models for predicting Revenue (from raw data or with these sketches themselves), and applying these models to identify slices that are anomalous. This is leveraged to answer questions such as what slices are relevant to look at today in O(n) time (usually milliseconds), comparing two different time periods or metrics that user picks interactively to identify slices that have changed abnormally and so on.

## Why does this work?

Most relevant analyses that look at slices only care about slices that impact the metric enough. If a slice is large enough to have an impact it gets captured by the sketching pass. This allows running the analysis on this materialized portion of the cube and leveraging the results as an approximation of the original analyses. While in theory some insights could be missed due to the approximations introduced, in practice this technique works really well and it was highly effective for our business partners with no known recall issues.

## Implementation - Efficiency and Accuracy

Most of our implementation was written in golang with reasonable optimizations. Some experiments with C++ provided a speed up on the offline computations, however we were able to absorb most of those into golang. We left a bit more on the table in exchange for speed of development for the rest of the systems.

Our initial implementation started with a single pass over the data to generate the appropriate Top n, however for higher dimensional data (num columns > 20), that started to slow things down particularly for a higher number of attribute combinations (3-4), which we call depth. Experimenting with multiple passes gave us speed improvements even while going through the data multiple times. The simplest optimization was to do a pass for each level that reduces the possibilities for deeper slices, e.g. for depth=n+1, only the ones that are "children" of depth=n slices are considered. So even with multiple passes, this turns out to be much better, particularly for examples with 100+ columns.

We experimented with various values for Top n and 100k seemed to fit most use cases well without loss of quality. In cases where the number of columns was high (say 100+), increasing this limit to 1M helped sometimes. The pruning pass to keep the Top n can be expensive, and increasing the n reduces the number of times pruning is required, thus improving overall speed.

For many cases, we decided to add another pass at the end, to get exact values for each of the slices we had picked, so instead of a range of values, we knew the exact metric value for these slices. This helped for metrics such as COUNT and SUM. DISTINCT COUNT or PERCENTILE still had to be approximated, but without a range.

For large numbers of columns, often many columns are related e.g. City, State, Country where City (or City, State) implies Country, or SubCategory implies Category. In cases like this, doing a first pass on a sample of the data to identify these relationships makes the eventual processing more efficient and removes duplicates from the Top n thus improving coverage as well.

## Types of Metrics

TRACE works for metric aggregations such as COUNT and SUM, combinations of multiple columns such as AVERAGE, RATIOs, SUMs and DIFFERENCEs, and even for aggregations such as DISTINCT COUNT ([Flajolet](), [Kane]()) and PERCENTILE. The only requirements are sketching algorithms that can approximate the metric with a reasonable accuracy, ideally with an algorithm where sketches support "addition" and "subtraction" since that enables more types of analysis such as the ones we use for identifying the most relevant ones.

### Composite Metrics

Metrics that are computed as a combination of two or more simple (base) metrics such as COUNT and SUM (or DIFF) are handled by maintaining the sketches for each base metric and combining them at run time. To ensure consistent results we maintain sketches for a union of slices that are relevant (Top n) for the base metrics.

Since the user cares about the value across all records, the slices that have a large denominator value (sum or count) end up making a bigger impact on that value than a slice that has a very high or low ratio value. Some ratios that we worked with: *conversion rates*, *click through rates*, *average order value*.

Metrics with positive and negative values are harder to deal with and one way they are handled in this system is to create separate base metrics for positive and negative values, and then combine them afterwards. This results in identification of slices that have Top n for |metric| which is more relevant for the use cases than purely Top n for metric.

## Online Analysis

With the Top n slices available for each time period, here are some of the analyses that can be done interactively:
- Compare slices across different time periods e.g. Revenue this week vs 3 months ago.
- Compare across slices e.g. how does revenue in Washington differ from that in California (proportion of sales).
- With a model that generates predictions (or "expected" values) for these slices, which slices are anomalous - these are usually the ones to look at.
- Comparing predictions from a model to actual values to identify gaps in model quality.

## Relevance

All these analyses (and others) that are done over a large number of slices have the problem of generating too many possible answers, typically involving a slice along with its parent(s) and children. For a slice such as [State=California, Device=Mobile], [State=California] and [Device=Mobile] are parents and [State=California, Device=Mobile, Category=Electronics] is an example child. Often in manual analysis this is worked around by starting from the top, and slowly going deeper into children as needed. This is ok, however it misses cases where children may be relevant and interesting but it gets hidden by the parent e.g. because other children create noise or cancel the effect out. Since our "cube" allows us to look at multiple levels simultaneously, this can be leveraged to identify the slice that is most relevant for a specific analysis. This is done by comparing it (and its change) to its parent(s) and child(ren) and leveraging causal analysis to pick the right one.

### Significant slices

With the cube containing n slices, not all are important all the time. For the case of "comparisons" and "change over time", some statistical significance techniques can be applied to remove noise. Since the cube contains relevant slices for each "time unit" the data across time can be used for "change over time" queries to filter out changes that are typical variations. With much of this data available online, optimized queries allowed us to do most of this computation online, thus keeping the querying phase separate from the cubing phase.

For other types of queries, there's also potential to leverage "population" based significance techniques to identify significant differences, beyond the change in mean.

### Drivers

Using a graph of all parent, child relationships that are available, leveraging counterfactuals of each slice allows us to identify likely drivers. This is particularly true for comparisons with "baselines" that is a common use case, although we were able to apply it to other use cases in identifying key leverage points for ratios such as conversion rates. The counterfactual here could be a predicted one (if a slice performed as expected) or a specific past value, and is used to identify impact on the top level to identify the most important slices. Prior work in this area such as [Cascading Analysts](), [Data X-Ray](), [Scorpion]() rely on hierarchical analysis which while relevant has a chance of missing real issues, whereas going over all *important* slices in our case allows the algorithms to be faster and provide better results.

## Related Work

[Macrobase]() focuses on optimizing each query to the data by leveraging sketching techniques and combining the explanation and classification pieces into the same pass. Our work differs in that it leverages a single pass to materialize most important slices, thus allowing rapid analysis of many different types over the materialized cube.

[DIFF]() builds on Macrobase by providing a difference across two specific time slices via a relational query interface. The materialization with TRACE allows a more interactive experience for DIFF like analysis, and allows for many others. It is orthogonal in that operators like DIFF could be applied on top of TRACE as well.

[HoCA]() focuses on a logical cubing framework allowing formulation of analytical queries like the ones we discuss, however it leaves out the computation cost and materialization of the cube as an independent task. The speed ups provided by TRACE are again orthogonal to the work from HoCA and enable many more interactive use cases that are unable to take advantage of the generality without the speed.

## Conclusion

In this paper we describe a Time-Relational Approximate Cubing Engine (TRACE) for speeding up analytical queries to make them interactive. This was achieved by picking out the most important parts of the cube to materialize, taking advantage of the fact that many user metrics have a clear notion of importance thus allowing us to ignore large parts of the possible space. Our implementation of TRACE was fast and useful for a variety of analysis across a range of business metrics and use cases.

We also experimented with providing the results from the cube via SQL that addresses the original table/dataset that was processed. This allows the cube to sit as a cache in-between and speed up typical queries from minutes to milliseconds. The results were promising however it would need more work to handle all cases and make it ready for common usage.

In addition to providing standard aggregations, the value of the cube is to provide online analysis and we experimented with SQL like options to enable data scientists to run custom analysis interactively on the cube. This is closer to the direction that HoCA takes, however instead of running on virtual cubes, we have a materialized cube that makes the process interactive.

## References

1. Firas Abuzaid, Peter Bailis, Jialin Ding, Edward Gan, Samuel Madden, Deepak Narayanan, Kexin Rong, and Sahaana Suri. Macrobase: Prioritizing attention in fast data. ACM Trans. Database Syst., 43(4):15:1–15:45, 2018.
2. Firas Abuzaid, Peter Kraft, Sahaana Suri, Edward Gan, Eric Xu, Atul Shenoy, Asvin Ananthanarayan, John Sheu, Erik Meijer, Xi Wu, Jeffrey F. Naughton, Peter Bailis, and Matei Zaharia. DIFF: a relational interface for large-scale data explanation. VLDB J., 30(1):45–70, 2021. doi: 10.1007/s00778-020-00633-6. URL https://doi.org/10.1007/s00778-020-00633-6.
3. Charikar, M., Chen, K., Farach-Colton, M. (2002). Finding Frequent Items in Data Streams. In: Widmayer, P., Eidenbenz, S., Triguero, F., Morales, R., Conejo, R., Hennessy, M. (eds) Automata, Languages and Programming. ICALP 2002. Lecture Notes in Computer Science, vol 2380. Springer, Berlin, Heidelberg. https://doi.org/10.1007/3-540-45465-9_59
4. DataSketches.org

5. Ronald Fagin, Ramanathan V. Guha, Ravi Kumar, Jasmine Novak, D. Sivakumar, and Andrew Tomkins. Multi-structural databases. In Chen Li (ed.), Proceedings of the Twenty-fourth ACM SIGACT-SIGMOD-SIGART Symposium on Principles of Database Systems, June 13-15, 2005, Baltimore, Maryland, USA, pp. 184–195. ACM, 2005a.
6. Philippe Flajolet, G. Nigel Martin, Probabilistic counting algorithms for data base applications, Journal of Computer and System Sciences, Volume 31, Issue 2, 1985, Pages 182-209, ISSN 0022-0000, https://doi.org/10.1016/0022-0000(85)90041-8
7. Kane, Daniel M., Jelani Nelson, and David P. Woodruff. 2010. "An Optimal Algorithm for the Distinct Elements Problem." In Proceedings of the Twenty-Ninth ACM SIGMOD-SIGACT-SIGART Symposium on Principles of Database Systems of Data: PODS '10, June 6-11, 2010, Indianapolis, Indiana: 41-52. New York, NY: ACM.
8. Matthias Ruhl, Mukund Sundararajan, and Qiqi Yan. The cascading analysts algorithm. In Gautam Das, Christopher M. Jermaine, and Philip A. Bernstein (eds.), Proceedings of the 2018 International Conference on Management of Data, SIGMOD Conference 2018, Houston, TX, USA, June 10-15, 2018, pp. 1083–1096. ACM, 2018. doi: 10.1145/3183713.3183745. URL https://doi.org/10.1145/3183713.3183745.
9. Xiaolan Wang, Xin Luna Dong, and Alexandra Meliou. Data x-ray: A diagnostic tool for data errors. In Timos K. Sellis, Susan B. Davidson, and Zachary G. Ives (eds.), Proceedings of the 2015 ACM SIGMOD International Conference on Management of Data, Melbourne, Victoria, Australia, May 31 - June 4, 2015, pp. 1231–1245. ACM, 2015.
10. Eugene Wu and Samuel Madden. Scorpion: Explaining away outliers in aggregate queries. Proc. VLDB Endow., 6(8):553–564, 2013.
11. Xi Wu, Shaleen Deep, Joe Benassi, Yaqi Zhang, Fengan Li, Uyeong Jang, James Foster, Stella Kim, Yujing Sun, Long Nguyen, Stratis Viglas, Somesh Jha, John Cieslewicz, and Jeffrey F. Naughton. Holistic cube analysis: A query framework for data insights, 2023. arXiv:2302.00120, URL https://doi.org/10.48550/arXiv.2302.00120